\documentclass[prd,preprint,tightenlines,floatfix,showpacs,preprintnumbers,nofootinbib,eqsecnum]{revtex4-1}

\pdfoutput=1

\usepackage[dvips,final]{graphicx}
\usepackage{color}
\usepackage{amssymb}
\usepackage{amsmath}
\usepackage{amsfonts}
\usepackage{epsfig}
\usepackage{bm}
\usepackage{comment}
\usepackage{float}
\usepackage[utf8]{inputenc}
\usepackage{caption}
\usepackage{subcaption}

\def\dfr{\mathrm{d}}

\DeclareSymbolFont{myletters}{OML}{ztmcm}{m}{it}
\DeclareMathSymbol{\uplambda}{\mathord}{myletters}{"15}

\begin{document}

\title{Momentum transfer squared dependence of exclusive quarkonia 
photoproduction in UPCs}

\author{Cheryl Henkels$^{1}$}
\email{cherylhenkels@hotmail.com}

\author{Emmanuel G. de Oliveira$^{1}$}
\email{emmanuel.de.oliveira@ufsc.br}

\author{Roman Pasechnik$^{1,2}$}
\email{Roman.Pasechnik@thep.lu.se}

\author{Haimon Trebien$^{1}$}
\email{haimontrebien@outlook.com}

\affiliation{
\\
{$^1$\sl Departamento de F\'isica, CFM, Universidade Federal 
de Santa Catarina, C.P. 476, CEP 88.040-900, Florian\'opolis, 
SC, Brazil
}\\
{$^2$\sl
Department of Astronomy and Theoretical Physics, Lund
University, SE-223 62 Lund, Sweden
}}

\begin{abstract}
\vspace{0.5cm}
In this paper, we study fully differential quarkonia photoproduction 
observables in ultraperipheral collisions (UPCs) as functions 
of momentum transfer squared. We employ the dipole picture of 
the QCD part of the scattering with proton and nucleus 
targets, with the projectile being a quasi-real photon flux 
emitted by an incoming hadron. We analyse
such observables for ground $J/\psi$, $\Upsilon(1S)$ and excited 
$\psi'$, $\Upsilon(2S)$ states whose light-front wave functions 
are obtained in the framework of interquark potential model
incorporating the Melosh spin transformation. Two different 
low-$x$ saturation models, one obtained by solving the 
Balitsky--Kovchegov equation with the collinearly improved kernel 
and the other with a Gaussian impact-parameter 
dependent profile, are used to estimate
the underlined theoretical uncertainties of our calculations.
The results for the proton target and with charmonium in the final state 
are in agreement with the available HERA data, while in the case 
of nucleus target we make predictions for $\gamma A$ and $AA$ differential 
cross sections at different $W$ and at $\sqrt{s}=5.02$ TeV, respectively.
\end{abstract}


\maketitle

\section{Introduction}
\label{Sect:intro}

The determination of the structure of protons and nuclei in terms of their fundamental constituents 
as well as their interactions is one of the biggest goals of particle physics \cite{Mantysaari:2020axf}. 
An important milestone for the proton structure measurements was the start of operation of the HERA collider 
at DESY. There, a large amount of Deep Inelastic Scattering (DIS) data (in which simple point-like leptons 
are used to probe the proton substructure) has been collected, making it possible to extract a detailed 
knowledge about the parton distribution functions (PDFs) for the proton with a good precision 
for as low longitudinal momentum fraction $x$ as $10^{-5}$ or so \cite{Glazov:2007zz}.

In order to obtain a more detailed picture of the target, in particular, to access an information about its 
transverse shape at a given $x$, more differential observables are needed. Two processes that provide such 
observables, the DVCS (where the outgoing photon is real) and the exclusive production of vector mesons (with the same 
quantum numbers $J^{PC} = 1^{--}$ as those of the photon), are frequently discussed in the literature. 
In the first case, thanks to the high 
beam energy available at the HERA collider, the experiments H1 and ZEUS have measured the pure DVCS 
cross section for the Bjorken variable ranging between $10^{-4}$ and $10^{-2}$. In the second case,
besides exclusive electro- and photoproduction of light vector mesons ($\phi$, $\rho$) and quarkonia ($J/\psi$)
studied by the H1 and ZEUS collaborations, there are more recent data on vector meson photoproduction 
in ultraperipheral collisions (UPCs) available from the LHC. The latter processes are in the main 
focus of this work.

Particle production processes in proton-nucleus $pA$ and nucleus-nucleus $AA$ UPCs
have attracted a lot of attention in recent years due to their vast potential in
probing the proton and nucleus structure at very small $x$ (for a recent review, 
see e.g. Ref.~\cite{Schafer:2020bnm}).
A particularly clean environment in UPCs is achieved in a fully exclusive process 
when a small-mass hadronic system is produced being separated from the intact 
scattered particles by large rapidity gaps on both sides. A phenomenologically important
and well-known example of such a scattering refers to exclusive quarkonia (such as charmonia 
$J/\psi\equiv \psi(1S)$, $\psi'\equiv \psi(2S)$ and bottomonia $\Upsilon(1S,2S)$)
photoproduction reactions in UPCs that has recently gained a particular relevance 
motivated by a wealth of experimental data coming from the LHC, such as those 
from LHCb \cite{Aaij:2015kea,LHCb:2018ofh,Aaij:2018arx}, ALICE \cite{Abelev:2012ba,Abbas:2013oua,Adam:2015sia,Kryshen:2017jfz,Acharya:2019vlb, Acharya:2021bnz} 
and CMS \cite{Khachatryan:2016qhq,Sirunyan:2018sav} experiments. 

The process is straightforwardly visualised by considering it in the target rest frame. 
While on one side of the collision, a photon flux is being emitted from a 
fast projectile (hadron or nucleus) and then fluctuate into a color-neutral $Q\bar Q$ ($Q=c,\,b$) 
pair called a color dipole, on another side such a dipole coherently rescatters off the target 
by means of an exchange of multiple gluonic system in a color-singlet state -- a dominating 
configuration at low longitudinal momentum transfers, $x$. In the leading-order perturbative 
Quantum Chromodynamics (QCD) approximation, typically validated by having a hard scale associated 
with the heavy-quark mass $m_Q$, one considers a colorless gluon-pair exchange between the dipole 
and the target. In the limit of small $x\ll 1$ and low four-momentum transfer squared 
$|t|=-(p_1-p_1')^2\ll m_Q^2$, such an exchange in momentum space is usually described 
in terms of the generalised unintegrated gluon density in the target which, in turn, 
connects to the dipole scattering matrix as a function of gluon $x$, dipole separation 
$\vec r$ and the impact parameter of the scattering $\vec b$. This matrix effectively 
encodes dynamics of parton saturation as well as contains full information about 
the relative dipole orientation with respect to the color background field of the target.
As long as $\vec r$ is integrated out in a convolution with the quarkonium light-front (LF) 
wave function, the impact parameter dependence provides the transverse profile 
of the target gluon density that can be probed by means of the measured differential 
in $t$ distributions. 

The impact-parameter dependence of the gluon density in the target is an intrinsically 
non-perturbative property and is often parameterised in terms of a Gaussian distribution 
like it is done, for example, in the case of the so-called ``bSat'' model 
\cite{Kowalski:2006hc}. In order to get a more accurate description of interactions 
between the color dipole and the target encoded in the impact-parameter profile of 
the target, the corresponding amplitude can be found by solving the Balitsky-Kovchegov (BK) 
evolution equation \cite{Balitsky:1995ub, Kovchegov:1999ua}. It is known that the BK equation at the next-to-leading order (NLO) 
is unstable due to large NLO corrections when one integrates out the gluon emissions 
with small transverse momenta. So, these corrections need to be properly resummed to 
all orders \cite{Ducloue:2019jmy}. Besides, an additional phenomenon called the Coulomb tails 
that corresponds to an unphysical growth of the amplitude at large impact parameters should
be taken into consideration. The latter phenomenon is found to be connected to the creation of large daughter dipoles during the evolution, thus enabling this problem to be cured. The BK solutions 
without such Coulomb tails can be found in several recent studies, 
e.g.\ in Refs.~\cite{Bendova:2019psy,Cepila:2020xol} this problem is absent by the use of the collinearly improved kernel. In the current analysis, we apply both
the ``bSat'' model and the BK solution with collinearly
improved kernel in the study of differential
quarkonia photoproduction cross sections in UPCs for relevant experimental conditions 
at HERA and LHC colliders.

The paper is organised as follows. In Sect.~\ref{sec:UPCs-proton}, we give a short 
description of the differential cross section of elastic vector meson photoproduction
$\gamma p \to V p$ off the proton target in terms of the dipole $S$-matrix and quarkonia 
LF wave functions in the framework of potential approach. In Sect.~\ref{Sec:dipole}, we
discuss the models for the impact-parameter dependent partial dipole amplitude that have
been used in the numerical analysis throughout this work. Sect.~\ref{Sec:results-proton}
presents the numerical results for the differential cross section of the 
$\gamma p \to V p$ process for the ground and excited quarkonia states, with $J/\psi$
results successfully describing the existing data. In Sect.~\ref{Sec:nuclear}, we review the
formalism to obtain the differential cross section 
of coherent quarkonia photoproduction off nuclear targets in UPCs and show our corresponding 
numerical predictions for the ground and first excited $\psi$ and $\Upsilon$ states
presented in Sect.~\ref{Sec:nuclear-results}. At last, a brief summary of our results 
is given in Sect.~\ref{Sect:conclusions}.

\section{Elastic photoproduction off a proton}
\label{sec:UPCs-proton}

The advantage of studying the vector meson photoproduction is that, in order to produce a single vector meson 
and nothing else in a detector, a color charge cannot be transferred to the target, requiring that at least two gluons (in the net color-singlet state) are exchanged. This provides an exclusive character of the process, with a particularly clean environment. Another advantage is that only in the exclusive scattering process it is possible to measure the total momentum transfer $\Delta_T$, and interpret it as the Fourier conjugate of the impact parameter (see e.g. Ref. \cite{doi:10.1142/S2010194517600588, Ducati:2017Na}). Consequently, these processes probe not only the density of partons, but also their spatial distribution in the transverse plane.

Considering first the proton target case, at high energies the elastic diffractive differential cross 
section for the $\gamma p \rightarrow V p$ scattering is found as follows \cite{Kowalski:2006hc}:
\begin{equation}
\frac{\mathrm{d} \sigma^{\gamma p \rightarrow V p}}{\mathrm{d} t}=
\frac{1}{16 \pi}\left|\mathcal{A}^{\gamma p}(x,\Delta_T)\right|^{2} \,,
\label{dif_cross_sec}
\end{equation}
where $t = -\Delta_T^2 \equiv (p_1-p_1')^2$ is the momentum transfer squared, $\Delta_T\equiv|\boldsymbol{\Delta}|$ 
is the transverse momentum of the produced vector meson $V$ recoiled against the target (assuming the projectile 
photon momentum to be collinear i.e.\ carries no transverse momentum), and the elastic production amplitude
\begin{eqnarray}
\mathcal{A}^{\gamma p}(x, \Delta_T)= \int \mathrm{d}^{2} 
\boldsymbol{r} \int_{0}^{1} \mathrm{d} z \, \left(\Psi_{V}^{*} \Psi_\gamma\right)\, 
\mathcal{A}_{q \bar{q}}(x, \boldsymbol{r}, \bm{\Delta}) \,,
\end{eqnarray}
is given in terms of the overlap between the transversely-polarised real 
photon $\gamma \to Q\bar Q$ ($\Psi_\gamma$) and vector meson $V \to Q\bar Q$ LF 
wave functions ($\Psi_\gamma$ and $\Psi_{V}$, respectively). 
Here, the elementary amplitude for elastic $q\bar q$ dipole scattering $\mathcal{A}_{q \bar{q}}$ 
is related to the dipole $S$-matrix 
\begin{equation}
\begin{aligned}
\mathcal{A}_{q \bar{q}}(x, \boldsymbol{r}, \bm{\Delta}) &=\int \mathrm{d}^{2} \boldsymbol{b} \,
\mathrm{e}^{-\mathrm{i} \boldsymbol{b} \cdot \boldsymbol{\Delta}} \, \mathcal{A}_{q \bar{q}}(x, \boldsymbol{r}, \boldsymbol{b}) = 
\mathrm{i} \int \mathrm{d}^{2} \boldsymbol{b} \, \mathrm{e}^{-\mathrm{i} \boldsymbol{b} 
\cdot \boldsymbol{\Delta}} \, 2[1-S(x, \boldsymbol{r}, \boldsymbol{b})] \,. \label{Aqq}
\end{aligned}
\end{equation}
and thus contains the most detailed (5-dimensional) information about the gluons density in the target.
It is directly connected to the so-called gluon Wigner distribution as was established earlier 
in Ref.~\cite{PhysRevLett.116.202301}. Even though a direct access of the elliptic gluon density in the Wigner distribution by a measurement of the exclusive quarkonia photoproduction is impossible, due $\boldsymbol{r}$ variable being integrated in the measured differential cross section, an access of the impact parameter profile of the target gluon density is still very relevant for understanding the hadron or nucleus structure at very low momentum transfers. 

Note, by means of the optical theorem, the imaginary part of the partial 
dipole amplitude in the forward limit ($\Delta_T\to 0$) is related to the dipole cross section 
$\sigma_{q\bar q}(x,r)$ -- a universal ingredient whose parameterization can be extracted 
from a given process (typically, from DIS) and then used for description of many 
other processes in $ep$, $pp$ and $pA$ collisions \cite{Kopeliovich:1981pz,Nikolaev:1994kk} 
(for a first analysis of elastic charmonia photoproduction in the dipole picture, see
e.g.~Refs.~\cite{Kopeliovich:1991pu,Kopeliovich:1993pw,Nemchik:1994fp,Nemchik:1996cw,Ducati:2013bya}).

In the off-forward case, one straightforwardly rewrites the elastic amplitude in terms
of the imaginary part of the elastic $q\bar q$ amplitude in the impact parameter representation 
in the following way \cite{Kowalski:2006hc}
\begin{equation}
\mathcal{A}^{\gamma p}(x, \Delta_T) = 2i \int \mathrm{d}^{2} \boldsymbol{r} 
\int_{0}^{1} \mathrm{d} z \int \mathrm{d}^{2} \boldsymbol{b}\left(\Psi_{V}^{*} \Psi\right)
\mathrm{e}^{-i[\boldsymbol{b}-(1-z) \boldsymbol{r}] \cdot \boldsymbol{\Delta}} 
N(x, \boldsymbol{r}, \boldsymbol{b}) \,.
\label{final_expr_amplitude}
\end{equation}
where $z$ is the longitudinal momentum fraction of a heavy (anti)quark in the $Q\bar Q$ dipole, and
\begin{eqnarray}
N(x, \boldsymbol{r}, \boldsymbol{b}) \equiv 
{\rm Im}\mathcal{A}_{q \bar{q}}(x, \boldsymbol{r}, \boldsymbol{b}) = 
2[1-{\rm Re}S(x, \boldsymbol{r}, \boldsymbol{b})] \,,
\end{eqnarray}
such that the dipole cross section is defined as follows,
\begin{eqnarray}
\sigma_{q\bar q}(x,r)=2\int \mathrm{d}^{2} \boldsymbol{b} \, N(x, \boldsymbol{r}, \boldsymbol{b}) \,.
\label{dipoleCS}
\end{eqnarray}

In order to take into account the real part of the $\mathcal{A}_{q \bar{q}}$ amplitude, 
it suffices to introduce in Eq.~(\ref{dif_cross_sec}) a factor that represents 
the ratio of the real to imaginary parts of the exclusive photoproduction amplitude 
$\mathcal{A}^{\gamma p}$ as follows \cite{Hufner:2000jb}:
\begin{equation}
    \mathcal{A}^{\gamma p} \Rightarrow  
    \mathcal{A}^{\gamma p} \left(1 - i \frac{\pi \lambda}{2} \right) \,, 
    \quad \mathrm{with} \quad
    \lambda = \frac{\partial \ln  \mathcal{A}^{\gamma p} }
    {\partial \ln (1/x)} \, . \label{lambda}
\end{equation}

At last, one typically also incorporates the so-called skewness effect of the off-diagonal gluon
distribution, which takes into account the fact that the gluons exchanged between the $q\bar{q}$ pair 
and the target can carry very different fractions of the target's momentum ($x$ and $x'$), while in 
the dipole cross section parameterisations fitted to the inclusive DIS data they appear 
to be same due to the optical theorem. So, considering the dominant kinematical configuration 
with $x' \ll x \ll 1$, the skewness effect is typically included via a multiplicative factor 
$R_g^2$ applied to the differential cross section in Eq.~(\ref{dif_cross_sec}) 
(see e.g.~Ref.~\cite{Shuvaev:1999ce}), with
\begin{equation}
R_g(\lambda)=\frac{2^{2 \lambda+3}}{\sqrt{\pi}} \frac{\Gamma(\lambda+5 / 2)}
{\Gamma(\lambda+4)} \,, \label{Rg}
\end{equation}
where $\lambda$ is found in Eq.~(\ref{lambda}).

Following our previous work \cite{Henkels:2020kju}, we have used the vector-meson wave functions calculated within the potential approach, which relies on factorisation of the wave function into the spin-dependent and radial components. In the rest frame of the color dipole, the radial wave function is found as a numerical solution of the Schr\"odinger equation, which can be solved for different models for the interquark potential and then boosted to the infinite momentum frame, where the dipole formula for the vector meson production amplitude (\ref{final_expr_amplitude}) is defined. In this analysis, we use five different models for the $Q\bar Q$  ($Q=c,b$) interaction potential: power-like model \cite{Martin:1980jx,Barik:1980ai} (pow), harmonic oscillator (osc), Cornell potential \cite{Eichten:1978tg,Eichten:1979ms} (cor), Buchm\"{u}ller-Tye parametrisation \cite{Buchmuller:1980su} (but) and logarithmic potential \cite{Quigg:1977dd} (log). These models have been fitted to the hadron spectrum and, when solving the Schr\"odinger equation, we have used the same parameters used in the original fits, including the heavy quark masses.

However, it is worth mentioning that the quark masses obtained by fitting the interquark potential are not bare masses, as they carry non-perturbative effects that are different for every potential and they are allowed to vary in order produce a better parametrization of the potential. So, in order to maintain the universality of the color dipole model, we chose to use, in the short-distance amplitudes, fixed perturbative masses given by $m_c=1.4$ GeV and $m_b=4.75$ GeV for charm and bottom quarks, respectively. 

When performing the Lorentz transformation between the two frames, not only the radial part should be properly boosted, but also the spin-dependent part has to be transformed accordingly. Such a transformation is known as the Melosh spin rotation of the quark spinors \cite{Melosh:1974cu} which causes an important impact on the differential photoproduction cross section, especially for excited quarkonia states \cite{Krelina:2018hmt,Cepila:2019skb} (for a detailed analysis of the Melosh spin rotation effect, see Refs.~\cite{Hufner:2000jb}). Indeed, the spin rotation increases the ground-state quarkonia cross sections by approximately 30\%, while for the excited states the increase is by a factor of 2-3 playing an important role in description of the exclusive vector meson photoproduction data.

Using such a quarkonium wave function in Eq.~(\ref{final_expr_amplitude}), the resulting photoproduction 
amplitude (considering the transversely-polarised real photon only) is given by
\begin{equation}
\begin{aligned}
\mathcal{A}_{T, L}^{\gamma p}(x, \Delta_T)=& 2i 
\int \mathrm{d}^{2} \boldsymbol{r} \int_{0}^{1} \mathrm{d} z \int \mathrm{d}^{2} \boldsymbol{b} \,
\mathrm{e}^{-i[\boldsymbol{b}-(1-z) \boldsymbol{r}] \cdot \boldsymbol{\Delta}} \\
&\times \left[ \Sigma^{(1)}(z, r) N(x, \boldsymbol{r}, \boldsymbol{b}) + 
\Sigma^{(2)}(z,r) N'_r(x, \boldsymbol{r}, \boldsymbol{b}) \right] \,,
\label{final_expr_amplitude_with_Melosh}
\end{aligned}\end{equation}
where $N'_r\equiv dN/dr$,
\begin{eqnarray}
&& \Sigma^{(1)} = \frac{Z_Q\sqrt{N_c\alpha_{\rm em}}}{2\pi\sqrt{2}}\,
2 K_0(m_Q r)\int \dfr p_T J_0(p_Tr)\Psi_{V}(z ,p_T)
p_T\,\frac{m_Tm_L+m_T^2-2z(1-z)p_T^2}{m_L+m_T} \,, \nonumber
\end{eqnarray}
and
\begin{eqnarray}
\Sigma^{(2)} = \frac{Z_Q\sqrt{N_c\alpha_{\rm em}}}{2\pi\sqrt{2}}\,
2K_0(m_Q r)\int \dfr p_T J_1(p_Tr)
\Psi_{V}(z ,p_T)\frac{p_T^2}{2}\,\frac{m_L+m_T+(1-2z )^2m_T}{m_T(m_L+m_T)} \,. 
\nonumber
\end{eqnarray}
Here, $\alpha_{\rm em}=1/137$ is the fine structure constant, $N_c=3$ is the number 
of colors in QCD, $Z_Q$ and $m_Q$ are the electric 
charge and the mass of the heavy quark, respectively, $J_{0,1}$ ($K_0$) are 
the (modified) Bessel functions of the first (second) kind, respectively, 
$p_T$ is the transverse momentum of the produced quarkonium state, and
\begin{eqnarray}
m_T = \sqrt{m_Q^2 + p_T^2} \,, \quad m_L = 2m_Q\,\sqrt{z(1-z)} \,.
\end{eqnarray}

It is worth to mention that there are still significant theoretical uncertainties in description of the vector meson wave functions. Besides the approach discussed above, there are also other attempts to model them. A very recent one \cite{Mantysaari:2021ryb} executes the calculations at the NLO level in $\gamma p$ collisions for longitudinally polarized photons making use of the CGC framework and proposing a wave function based upon NRQCD matrix elements \cite{Lappi:2020ufv}. Other study \cite{Sambasivam:2019gdd} modifies the dipole cross section to enhance the suppression of dipoles with large separations beyond the confinement length-scale (a correction important for small $Q^2$). The analysis of Ref.~\cite{Goncalves:2020vdp} is very similar to ours except that the boosted Gaussian has been utilized there to construct the vector-meson wave functions.

\section{Partial dipole amplitude}
\label{Sec:dipole}

For the main purpose of scanning of the impact-parameter profile of the target nucleon or nucleus,
we need an impact-parameter dependent (or ${\bf b}$-unintegrated) dipole cross section that can be
found in terms of the dipole $S$-matrix introduced in Eq.~(\ref{Aqq}). First, we tested seven different 
models available from the literature, and then we selected the two that best describe the exclusive 
vector meson photoproduction data from the HERA collider, namely, the impact parameter dipole 
saturation model \cite{Kowalski:2006hc} (dubbed as ``bSat'' in what follows) and the model 
based upon a numerical solution of the Balitsky-Kovchegov (BK) equation \cite{Bendova:2019psy}.

In the first case of ``bSat'', we employ the following formula
\begin{equation}
N(x, \boldsymbol{r}, \boldsymbol{b}) = 
1-\exp \left(-\frac{\pi^{2}}{2 N_c}\,r^2
\alpha_s(\mu^{2})\, xg(x, \mu^{2}) T(b)\right) \,,
\end{equation}
where $\mu^2 = 4/r^2 + \mu_0^2$ is the momentum scale in the collinear gluon
density $xg(x,\mu^2)$, and no non-trivial information about the relative dipole 
orientation is implemented. In numerical calculations, we have used 
the CT14LO parameterisation \cite{Dulat:2015mca} motivated by our earlier analysis
of integrated quarkonia photoproduction cross sections performed in Ref.~\cite{Henkels:2020kju}. This will be different from the original bSat model in which the gluon PDF is evolved up to the scale $\mu^2$ with LO DGLAP gluon evolution neglecting its coupling to quarks, but the numerical results will be similar enough to neglect the difference. 
Besides, we considered a conventional Gaussian form for the proton shape function $T(b)$
\begin{equation}
T(b) = \frac{1}{2\pi B_{\rm G}}\, 
\mathrm{e}^{-b^2/2B_{\rm G}} \,,
\end{equation}
where the slope parameter $B_{\rm G}=4.25\, \mathrm{GeV}^{-2}$ 
is found at Ref.~\cite{Kowalski:2003hm}.

In the second case, the numerical solution of the BK equation is provided 
by Ref.~\cite{Bendova:2019psy}, where it is obtained under the assumption 
that the dipole partial amplitude depends only on the absolute values of 
the transverse separation of the dipole $r$ and the impact parameter $b$, 
but does not depend on the angle between $\boldsymbol{r}$ and $\boldsymbol{b}$ 
similarly to the ``bSat'' model. In this case, the BK equation reads
\begin{equation}
\begin{aligned}
\frac{\partial {\cal N}(r, b, Y)}{\partial Y} = & \int d^2\boldsymbol{r}_{1} K(r, r_{1}, r_{2})
\Big( {\cal N}(r_{1}, b_{1}, Y) + {\cal N}(r_{2}, b_{2}, Y) - {\cal N}(r, b, Y) \\
& - {\cal N}(r_{1}, b_{1}, Y) {\cal N}(r_{2}, b_{2}, Y) \Big) \,
\end{aligned}
\end{equation}
whose numerical solution provides us with the partial dipole amplitude
\begin{equation}
    N(x, \boldsymbol{r}, \boldsymbol{b}) = {\cal N}(r,b,\ln(0.008/x))
\end{equation}
that has been employed in our numerical analysis below. The specific main feature of Ref.~\cite{Bendova:2019psy} solution is that it is obtained with a collinearly improved kernel $K(r, r_{1}, r_{2})$ studied in Ref.~\cite{Iancu:2015joa} that suppresses the larger daughter dipole sizes during the evolution and thus does not show the nonphysical Coulomb tails.

Finally, following Refs.~\cite{Drell:1969km, Brodsky:1973kr, West:1970av}, we also 
incorporate a correction relevant at large-$x$ multiplying the dipole cross section by 
a factor $(1-x)^{2 n_s - 1}$, where $n_s$ denotes the number of spectator quarks, which 
was chosen to be $n_s = 4$.

\section{Results for $\gamma p \rightarrow V p$ process}
\label{Sec:results-proton}

Now, that we have outlined the basic dipole formalism needed for analysis of the differential
photoproduction observables, let us first present the numerical results for the $\gamma p 
\rightarrow V p$ process. Note, in general the differential photoproduction cross sections 
computed for the proton target are very sensitive to the dipole parametrization used in the analysis. 
In this work, we analysed many different $b$-dependent parameterisations for the partial dipole amplitude, 
and they all give very different results. We chose to present the results obtained only with the BK solution 
and the ``bSat'' model briefly described above as those that provide the best description of 
the available $J/\psi$ data. We will start with the BK solution model.

\begin{figure}[H]
\begin{minipage}{0.48\textwidth}
 \centerline{\includegraphics[width=1.0\textwidth]{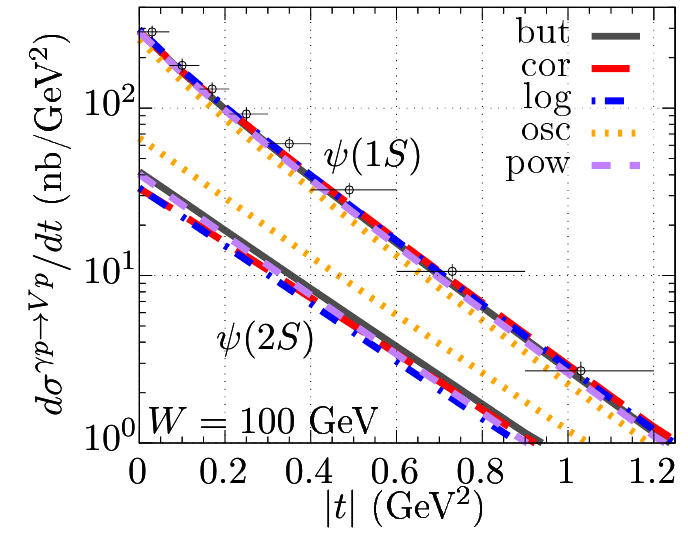}}
\end{minipage} \hfill
\begin{minipage}{0.48\textwidth}
 \centerline{\includegraphics[width=1.0\textwidth]{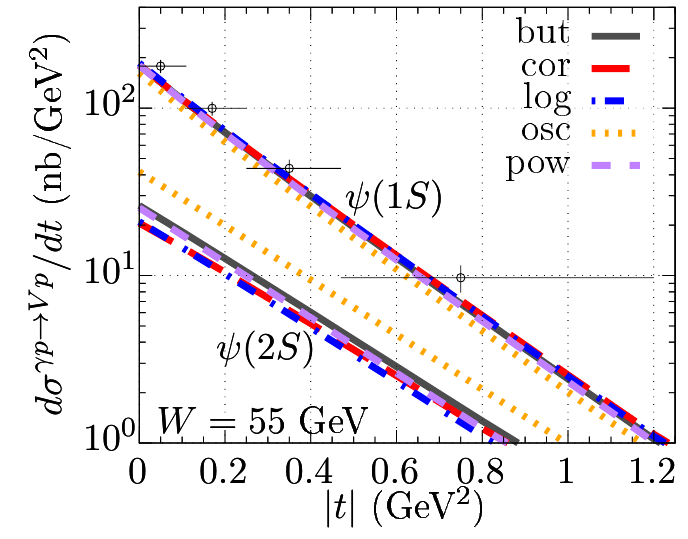}}
\end{minipage} \hfill
\caption{Differential cross section for $\psi(1S)$ (upper curves) and $\psi(2S)$ (lower curves) photoproduction 
as a function of $|t|$ obtained using the numerical solution of the BK equation  obtained in Ref.~\cite{Bendova:2019psy}, for $W = 100$ GeV (left) and $W = 55$ GeV (right). The results are presented for five different 
interquark potential models. The $\psi(1S)$ results are compared to the corresponding data from 
H1 Collaboration~\cite{Alexa:2013xxa,Aktas:2005xu}.
}
\label{gammap_cross_section_BK_psi}
\end{figure}

Fig.~\ref{gammap_cross_section_BK_psi} shows the differential cross section for $J/\psi\equiv \psi(1S)$ 
(upper curves) and $\psi(2S)$ (lower curves) production as a function of the momentum transfer squared 
$|t|$ for $W = 100$ GeV (left) and $W = 55$ GeV (right). 
Here, the results are obtained using a numerical solution of the BK equation of the 
$b$-dependent partial dipole amplitude discussed above. The ground-state charmonium results were compared to 
the experimental data available from the H1 Collaboration~\cite{Alexa:2013xxa,Aktas:2005xu} 
yielding a very good description. The corresponding observables have been evaluated
with the LF quarkonia wave functions obtained for several different parametrizations of 
the interquark $Q \bar{Q}$ potential (for more details, see Refs.~\cite{Cepila:2019skb, Henkels:2020kju}) 
which lead to a rather minor variation in the final results. A bigger 
difference is found for the $\psi(2S)$ cross section computed with the harmonic oscillator 
potential which is noticeably higher than the results for other potentials. This effect is due
to a specific shape of this wave function as was briefly discussed in Ref.~\cite{Henkels:2020kju}.
The $|t|$-slope is close to a constant due to an almost exponential impact parameter profile 
of the partial dipole amplitude, in full consistency with the $J/\psi$ data. One notices however 
a somewhat larger difference in the slopes of $J/\psi$ and $\psi(2S)$ differential cross sections
due different shapes of the wave functions.
\begin{figure}[!h]
\begin{minipage}{0.48\textwidth}
 \centerline{\includegraphics[width=1.0\textwidth]{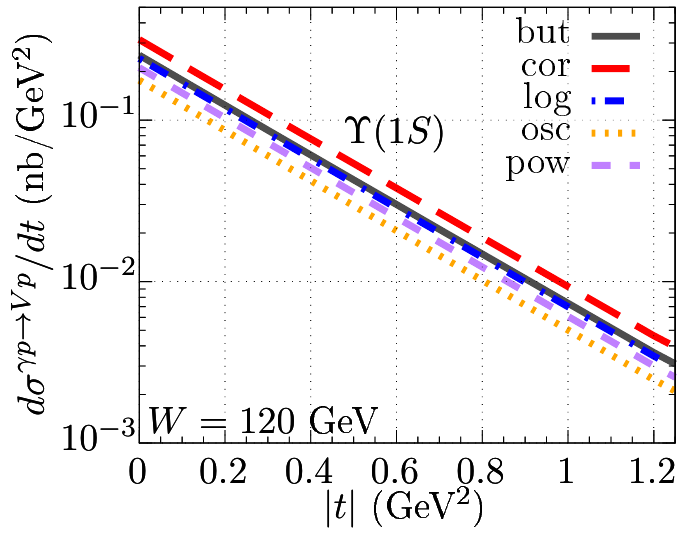}}
\end{minipage} \hfill
\begin{minipage}{0.48\textwidth}
 \centerline{\includegraphics[width=1.0\textwidth]{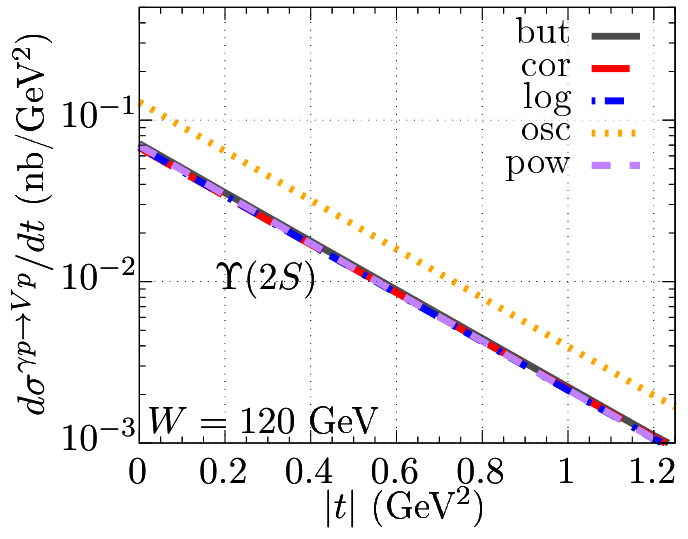}}
\end{minipage}\hfill
\caption{Predictions for the differential cross section for $\Upsilon(1S)$ (left) 
and $\Upsilon(2S)$ (right) as a function of $|t|$ obtained using the numerical solution of the BK equation obtained in Ref.~\cite{Bendova:2019psy}, for $W = 120$ GeV. The results are presented 
for five different interquark potential models.
}
\label{gammap_cross_section_BK_ups}
\end{figure}

In Fig.~\ref{gammap_cross_section_BK_ups} we present our predictions for the differential cross section 
of $\Upsilon(1S)$ (left) and $\Upsilon(2S)$ (right) photoproduction as a function of $|t|$, also using the numerical 
solution of the BK equation, for $W = 120$ GeV. The results for the ground and excited states are separated 
into two different plots since the corresponding results for the oscillator potential are very close. 
This occurs due to the fact that these two wave functions in the case of harmonic oscillator have 
a very similar small-$r$ dependence. Since this domain plays a dominant role in the integration 
of the $\Upsilon$ production amplitudes, one indeed arrives at very similar numerical results 
for $\Upsilon(1S)$ and $\Upsilon(2S)$ photoproduction in this case.
\begin{figure}[!h]
\begin{minipage}{0.48\textwidth}
 \centerline{\includegraphics[width=1.0\textwidth]{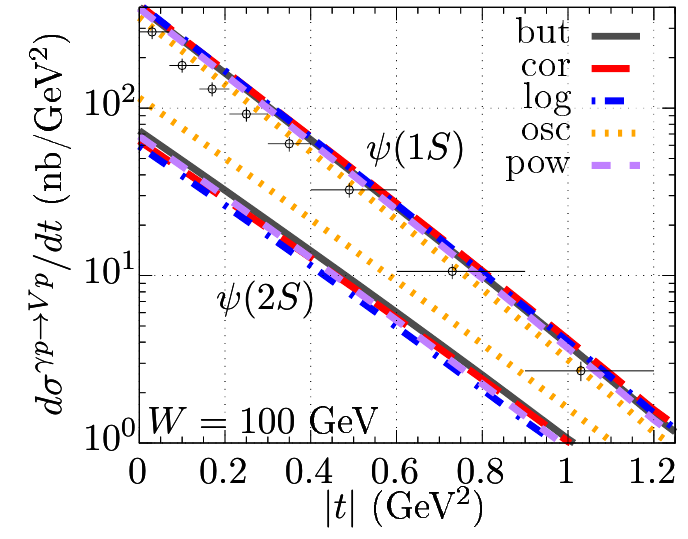}}
\end{minipage} \hfill
\begin{minipage}{0.48\textwidth}
 \centerline{\includegraphics[width=1.0\textwidth]{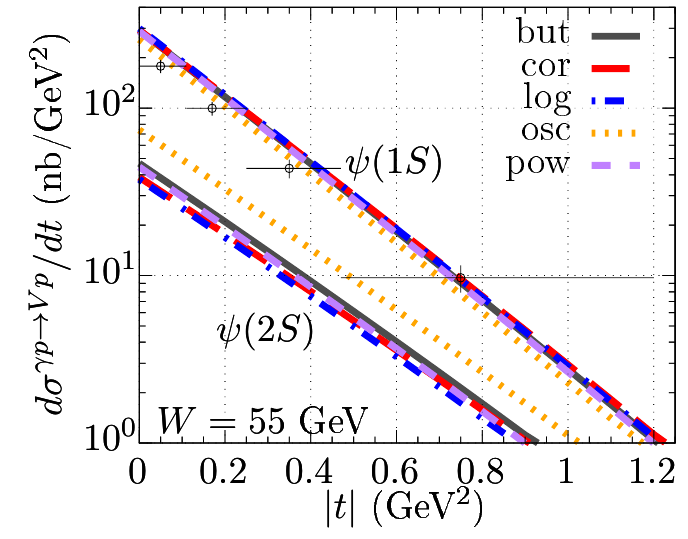}}
\end{minipage}
\caption{Differential cross section for $\psi(1S)$ (upper curves) and $\psi(2S)$ (lower curves) photoproduction
as a function of $|t|$ found with the the ``bSat'' dipole model for $W = 100$ GeV (left) 
and $W = 55$ GeV (right), including also the skewness effect. The results are presented 
for five different interquark potential models. The $\psi(1S)$ results are compared 
to the corresponding data from H1 Collaboration~\cite{Alexa:2013xxa,Aktas:2005xu}.
}
\label{gammap_cross_section_bsat_psi}
\end{figure}

\begin{figure}[!h]
\begin{minipage}{0.48\textwidth}
 \centerline{\includegraphics[width=1.0\textwidth]{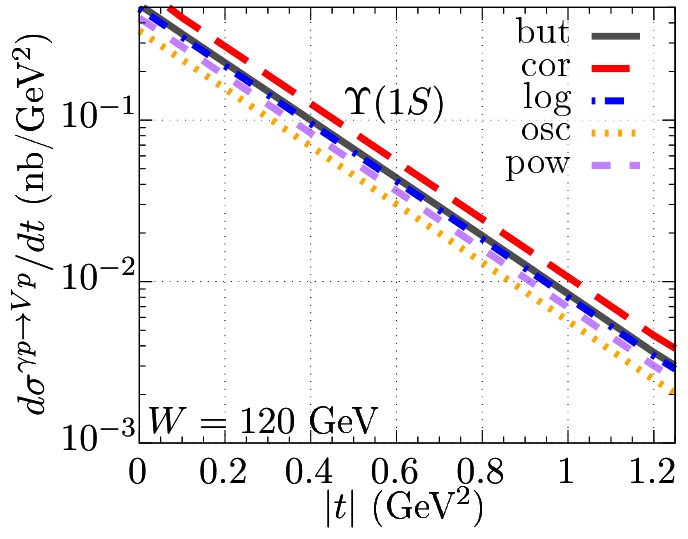}}
\end{minipage} \hfill
\begin{minipage}{0.48\textwidth}
 \centerline{\includegraphics[width=1.0\textwidth]{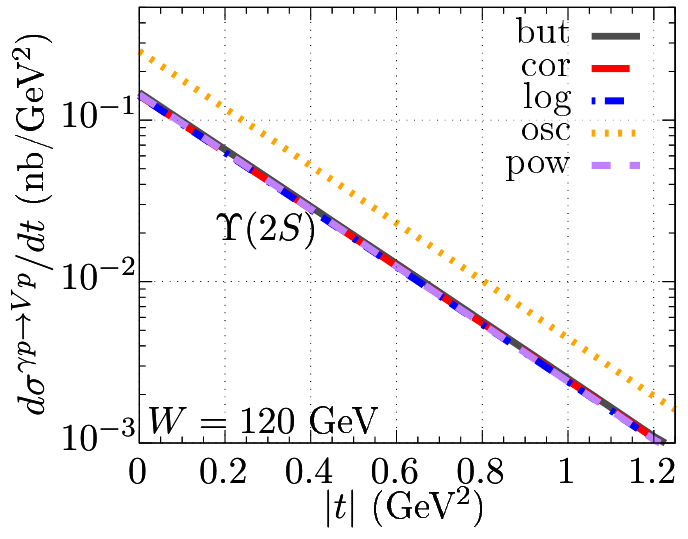}}
\end{minipage}
\caption{Predictions for the differential cross section for $\Upsilon(1S)$ (left) and 
$\Upsilon(2S)$ (right) photoproduction as a function of $|t|$ computed with the ``bSat'' dipole model 
for $W = 120$ GeV. The results are presented for five different interquark potential models.
}
\label{gammap_cross_section_bsat_ups}
\end{figure}

\begin{figure}[!h]
\begin{minipage}{0.48\textwidth}
 \centerline{\includegraphics[width=1.0\textwidth]{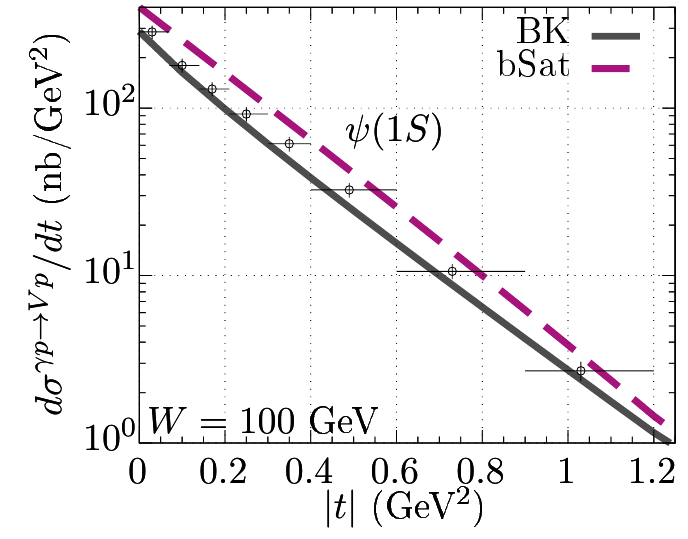}}
\end{minipage} \hfill
\begin{minipage}{0.48\textwidth}
 \centerline{\includegraphics[width=1.0\textwidth]{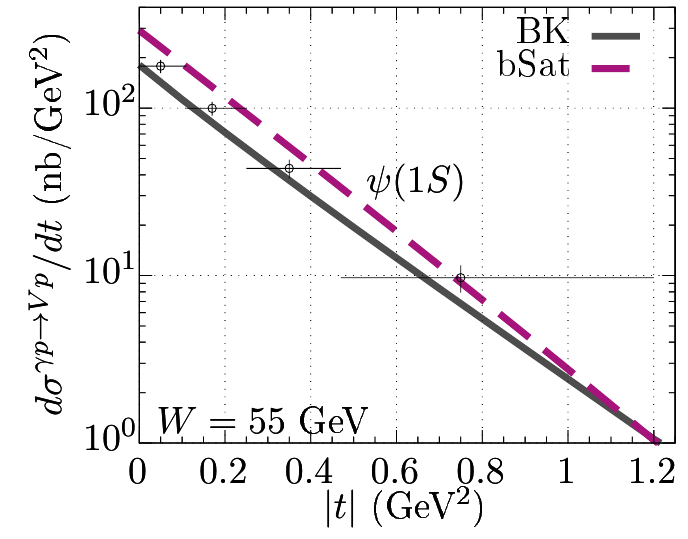}}
\end{minipage}
\caption{Differential cross section for $\psi(1S)$ photoproduction as a function of $|t|$ found using 
the Buchm\"{u}ller-Tye potential as well as the BK and ``bSat'' models for $W = 100$ GeV (left) 
and $W = 55$ GeV (right). The $\psi(1S)$ results are compared 
to the corresponding data from H1 Collaboration~\cite{Alexa:2013xxa, Aktas:2005xu}.
}
\label{gammap_cross_section_comp_BK-bsat}
\end{figure}

Figs.~\ref{gammap_cross_section_bsat_psi} and \ref{gammap_cross_section_bsat_ups} represent the same quantities 
as in Figs.~\ref{gammap_cross_section_BK_psi} and \ref{gammap_cross_section_BK_ups}, respectively, except that 
the former are computed with the ``bSat'' dipole parameterisation instead of the BK solution employed in the latter. 
As can be seen in Fig.~\ref{gammap_cross_section_bsat_psi}, the use of the ``bSat'' dipole model and 
the LF quarkonia wave functions calculated within the potential approach also provides a fair description 
of the H1 data. The latter is not as good as in the case of the BK solution though. However, since ``bSat'' dipole
parameterisation is widely used in the literature, in this work we chose to show the corresponding numerical 
results as well. A comparison between the curves obtained with these two dipole models and the available 
H1 data for $\psi(1S)$ photoproduction is presented in Fig.~\ref{gammap_cross_section_comp_BK-bsat}, 
where we can see that both curves found are mainly located within the experimental error bars for both $W = 100$ GeV (left) 
and $W = 55$ GeV (right), except that at small $|t|$ and at large $W$ the ``bSat'' model marginally overshoots 
the data.

\begin{figure}[!h]
	\centerline{\includegraphics[width=.5\textwidth]{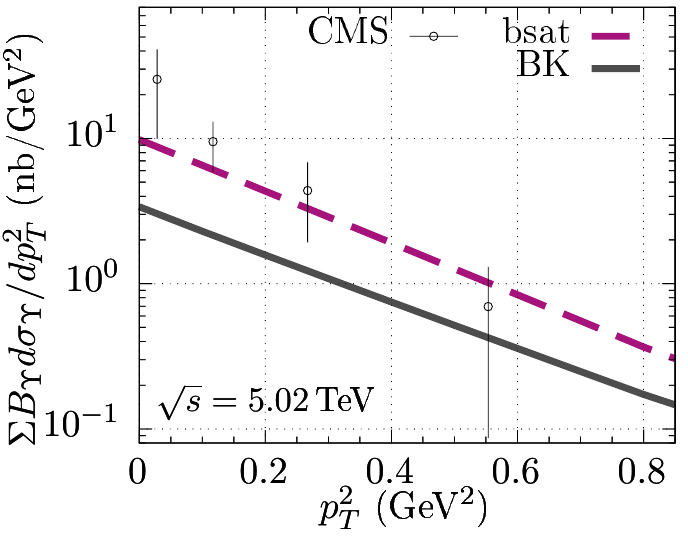}}
	\caption{Differential cross section for the $p Pb \rightarrow \Upsilon(nS) p Pb$ 
		process via $\gamma p \rightarrow \Upsilon(nS) p$  as functions of $p_T^2$, calculated with the Buchm\"{u}ller-Tye potential. The results using the BK and bSat dipole amplitudes are compared with the CMS data \cite{Sirunyan:2018sav}.}
	\label{pPb_cross_section} 
\end{figure}

At last, we include Fig.~\ref{pPb_cross_section} showing our results on the photoproduction cross section of $\Upsilon$ states in $p Pb$ collisions when the photon is emitted from the nucleus. The required photon flux will be discussed in the next section. These results are compared with the CMS data points \cite{Sirunyan:2018sav}. In the plot the curve is obtained by summing each $\Upsilon (nS)$ state contribution multiplied by its branching fraction in the dimuon decay channel, $B_{\Upsilon(nS)}$, taken from Ref.~\cite{ParticleDataGroup:2016lqr}. We notice that the models underestimate the $\Upsilon$ photoproduction data at small $p_T^2$, while getting closer to the data points at larger $p_T^2$. 
This shows that a fully satisfactory description of the data using the dipole model with an impact parameter dependence, in particular at large impact parameters, is still missing in the literature.

\section{Coherent photoproduction off nuclear targets}
\label{Sec:nuclear}

In photon-nucleus scattering, the differential cross section for coherent quarkonia 
$V$ photoproduction $\gamma A\to V A$ off a nuclear target with atomic mass $A$ can be 
found as follows:
\begin{equation}
    \frac{d \sigma^{\gamma A\to V A}}{dt} = \frac{1}{16 \pi} \,
    |\langle \mathcal{A}^{\gamma A}(x, \Delta_T) \rangle_N|^2 \,,
\end{equation}
in terms of the averaged amplitude \cite{Ivanov:2002kc}
\begin{equation}
\begin{split}
    \langle \mathcal{A}^{\gamma A}\rangle_N =&  2i\, \int \dfr^2 \textbf{r} 
    \int_0^1 \dfr z \int \dfr^2 \textbf{b}  e^{- i [\textbf{b} - (1-z)\textbf{r}] \cdot 
    \boldsymbol{\Delta}}\; \Sigma_T \, \left\langle N_A(x, \boldsymbol{r}, \boldsymbol{b}) \right\rangle_N \,, \label{averaged-A}
\end{split}
\end{equation}
where $\Sigma_T = \Sigma^{(1)} + \Sigma^{(2)} \partial/\partial r$, with the coefficients found in Eq.~(\ref{final_expr_amplitude_with_Melosh}). 
Following 
Ref.~\cite{Kowalski:2003hm}, the dipole-nucleus scattering amplitude averaged over 
all possible configurations of the nucleons in the target nucleus reads
\begin{equation}
    \left\langle N_A(x, \boldsymbol{r}, \boldsymbol{b}) \right\rangle_N = 
    1 - \left(1 - \frac{T_A(b) \sigma_{q\bar{q}}(x,r)}{2A} \right)^A \,. 
    \label{eq:averaged_amplitude}
\end{equation}
This equation was obtained using a $b$ dependent dipole amplitude parametrization, in the same way as above. 
It differs from other approach found in Ref.~\cite{Xie:2017mil}, where a Gaussian shape was assumed to describe
such $b$ dependence. The functions that appear in Eq.~(\ref{eq:averaged_amplitude}) are the usual (integrated) 
dipole cross section off the proton target, $\sigma_{q\bar{q}}(x,r)$, found in Eq.~(\ref{dipoleCS}), and
\begin{equation}
    T_A(b) = \int_{-\infty}^{+\infty} \dfr z \, 
\rho_A(b,z) \,, \qquad \frac{1}{A} \int \dfr^2 b \, T_A(b) = 1 \,,
\end{equation} 
i.e., the thickness function of the nucleus, given in terms 
of the normalised Woods-Saxon distribution \cite{Woods:1954zz},
\begin{eqnarray}
\rho_A(b,z) = \frac{{\cal N}}{1 + \exp[\frac{r(b,z) - c}{\delta}]} \,, \qquad
r(b,z)=\sqrt{b^2 + z^2} \,.
\end{eqnarray}
Here, $r(b,z)$ is the distance from the center of the nucleus and ${\cal N}$ 
is an appropriate normalisation factor. In this work, we consider UPCs of 
lead nuclei, with $A = 208$ and the parameters $c = 6.62$ fm and 
$\delta = 0.546$ fm are used~\cite{Euteneuer:1978qw}.

The amplitude in Eq.~(\ref{averaged-A}) takes into account the imaginary part of the amplitude only. 
In order to incorporate the real part, one performs the following substitution,
\begin{equation}
         \sigma_{q\bar{q}}(x,r) \Rightarrow \sigma_{q\bar{q}}(x,r) 
         \left(1 - i \frac{\pi \uplambda}{2}  \right) \quad \mathrm{ with} \quad
         \uplambda = \frac{\partial \ln \sigma_{q\bar{q}}(x,r) }
     {\partial \ln (1/x)} \,, \label{lambda-nucl}
\end{equation}
analogous to the one made for the proton target case in Eq.~(\ref{lambda}). Furthermore, 
in order to introduce a skewness correction to the associated nuclear gluon density, 
one can multiply the dipole cross section by the corresponding skewness factor 
found in Eq.~(\ref{Rg}) as
\begin{equation}
\sigma_{q\bar{q}}(r,x) \rightarrow \sigma_{q\bar{q}}(r,x) R_g(\uplambda) \,.
\end{equation}

In order to study the rapidity distribution of the vector mesons produced in $AA$ UPCs, 
one needs to incorporate the incoming photon flux $n(\omega)$ in one of the incident nuclei 
such that
\begin{equation}
    \frac{d \sigma^{A A \rightarrow V AA}}{dy dt} = n(\omega) 
    \frac{d \sigma^{\gamma A \rightarrow V A}}{dt}(y) + 
    \left\{ y \rightarrow -y \right\} \,,
    \label{rapidity_distribution}
\end{equation}
where $\omega  = (M_V/2) e^{y}$ is the projectile photon energy in the center-of-mass (c.m.) of the colliding 
particles given in terms of the mass of the vector meson $M_V$ and its rapidity $y$. The photon 
flux can be written as \cite{vonWeizsacker:1934nji,Williams:1934ad}
\begin{equation}
    n(\omega) = \frac{2 Z_A^2 \alpha_{\rm em}}{\pi} \left\{ \xi K_1(\xi )\, K_0(\xi) 
    - \frac{\xi^2}{2} \left[K_1^2(\xi) - K_0^2(\xi) \right] \right\} \,,
    \label{photon_flux}
\end{equation}
where $K_{0,1}$ are the modified Bessel functions of the second kind, $Z_A$ is the charge of the 
projectile nucleus sourcing the photon flux, $\xi = 2\omega R_A/\gamma$, 
$R_A$ is the radius of the nucleus (in the numerical analysis below we used $R_A$ value from 
Ref.~\cite{DeJager:1987qc}), $\gamma = \sqrt{s}/2m_p$ is the Lorentz factor, and $m_p$ is the proton mass.

This photon flux is also used in calculations of the cross section as a function of transverse momenta in $pA$ UPCs. The latter is computed in a similar way as Eq.~(\ref{rapidity_distribution}), except that we multiply the $\gamma p$ differential cross section by the photon flux (\ref{photon_flux}) in the first term, whereas the $\gamma A$ differential cross section is multiplied by the proton flux in the second term. At very small $t$ the $\gamma A$ (photon from proton) contribution is relevant, however, for not so small $t$, it falls much quicker than the $\gamma p$ (photon from nucleus) contribution and therefore this second term can be disregarded safely.

The gluon density inside a nucleus at small $x$ is expected to be suppressed compared to 
the one inside a free nucleon caused by a relative reduction of the dipole cross section due to 
interferences between incoming dipoles in the presence of the higher Fock states of the photon (see e.g. Refs.~\cite{Kopeliovich:1991pu,Ivanov:2002eq,Ivanov:2002kc} for more details).
This phenomenon also known as the nuclear (or gluon) shadowing effectively reduces 
the quarkonia photoproduction 
$\sigma^{\gamma A \rightarrow V A}$ cross section off a heavy nuclear target in comparison to that off 
the proton, $A\sigma^{\gamma p \rightarrow V p}$. Such a shadowing effect plays the most important role
at central rapidities of the meson, and can be phenomenologically incorporated by ``renormalising'' 
the dipole cross section as
\begin{equation}
\sigma_{q\bar{q}}(x,r) \rightarrow \sigma_{q\bar{q}}(x,r) R_G(x, \mu^2) \,,
\end{equation}
where $R_G$ is given in terms of a ratio of the gluon density function inside the heavy 
nucleus $xg_A(x,\mu^2)$ over the one inside the proton $xg_p(x,\mu^2)$ as
\begin{equation}
R_G(x, \mu^2) = \frac{xg_A(x,\mu^2)}{A\, xg_p(x,\mu^2)} \,.
\end{equation}
There is still a large uncertainty in the determination of the $R_G$ factor, and we highlight the past work of Refs.~\cite{PhysRevD.80.094004, PhysRevD.85.074028} for a thorough discussion of this issue. In our calculations, we employ the EPPS16 parameterisation for the nuclear 
gluon distribution fitted to the LHC data \cite{Eskola:2016oht}, adopting $\mu = M_V/2$ 
as the factorisation scale \cite{Guzey:2016piu}. As discussed in our previous paper~\cite{Henkels:2020kju}, the EPPS16 parametrisation was the one that has provided a more satisfactory description of the data using the standard factorisation scale; this fact motivates us to repeat this choice.

Besides the nuclear shadowing effect, there is another important correction to the coherent 
photoproduction cross section off a nucleus that is worth to be mentioned. In order to obtain 
the equations above, we used the Glauber-Gribov approach, which takes into account that the inelastic 
interactions with the nucleons in the target nucleus can produce particles that shortly thereafter 
can be absorbed by another bound nucleon effectively making the nucleus more transparent. These 
inelastic corrections are calculated considering that at high energies the dipole is an eigenstate 
of interaction, with its transverse separation being ``frozen'' in the course of its propagation 
through the target nucleus \cite{Kopeliovich:2005us, Kopeliovich:2016jjx}. This is called the ``frozen'' approximation and guarantees 
that there are no fluctuations of the $q \bar{q}$ dipole inside the nucleus. This approximation 
is only valid if the lifetime of the $q \bar{q}$ state, or the so-called coherence length,
\begin{equation}
    l_c = \frac{2 \nu}{M_V^2} \, ,
\end{equation}
is much larger than the nuclear radius i.e.\ $l_c \gg R_A$. Here, $\nu$ is the energy of the photon 
in the nucleus rest frame.

In the case where the coherence length is finite (i.e.~when it is not much larger than the nuclear 
radius ($l_c \lesssim R_A$)), one needs to incorporate additional corrections to the differential 
cross section $d\sigma^{\gamma A \rightarrow VA}/dt$, which depend on the c.m.\ energy $W$. 
This effect occurs because the photon can propagate through the nucleus without experiencing 
any attenuation until the $Q\bar{Q}$ fluctuation is produced. This propagation through 
the nucleus can be described mathematically by a light-cone Green function that satisfies 
a two-dimensional equation of motion (for more details, see Ref.~\cite{Kopeliovich:2001xj}), whose solution is 
known only for the quadratic dependence of the dipole cross section approximation $\sigma_{q\bar{q}} \propto r^2$ 
and for the oscillator form of the interquark potential. In our previous work \cite{Henkels:2020kju}, we used 
a simplified way to take this effect into account by multiplying the nuclear cross sections in the infinite 
coherence length limit by a form factor that can be found in Ref.~\cite{Ivanov:2002kc}. It has been shown
in a recent work of Ref.~\cite{Kopeliovich:2020has} that such an estimate is valid with a reasonable accuracy 
only for photoproduction of $\rho$ mesons. So the authors compared the vector dominance model 
with the approach based on the light-cone Green function and showed that there is a substantial 
difference between the form factors computed within each approach for small values of energy, 
mainly for the incoherent case. 

The effect of the finite coherence length is known to be sizable only at large values of rapidity, 
where there are not many measured data points. In this work we are focused on making 
the predictions for the differential quarkonia photoproduction cross sections at the LHC energies, 
and we chose to evaluate all the results for $y=0$. At this value, the finite coherence length effect 
does not affect the cross section so it can be safely disregarded.

\section{Results for $\gamma A$ and $AA$ collisions}
\label{Sec:nuclear-results}

Now, we would like to present the numerical results for the differential cross sections of coherent
vector meson production in $\gamma A$ collisions as well as in $AA$ UPCs. As was described above,
in our numerical calculations we employ the potential approach for $1S$ and $2S$ charmonia and bottomonia
LF wave functions incorporating the Melosh spin rotation. We chose to show the results obtained with the Buchm\"{u}ller-Tye potential since the difference from other potentials appear to be not significant for this analysis. The nuclear target dipole cross section will be built on top of three dipole--proton cross sections: the numerical solution of the BK equation, the phenomenological GBW model and also the bSat model. Furthermore, the effects 
of the nuclear shadowing have been accounted for using a phenomenological approach fitted to data \cite{Eskola:2016oht}.

In an attempt to improve the color dipole models available in the literature, the authors of Ref.~\cite{Kopeliovich:2021dgx} included a correlation between the impact parameter and the dipole separation in the elastic dipole-proton  amplitude, by adding a dependence on the angle between $\vec{r}$ and $\vec{b}$ to the color dipole cross section. It is known indeed that the interaction is amplified when $\vec{r}$ is parallel to $\vec{b}$ and vanishes when $\vec{r} \perp \vec{b}$. However, since there is an integration over all possible angles in the transverse plane, the inclusion of this correlation is expected to not make a significant impact to the final results presented in this paper.

\begin{figure}[!h]
    \begin{minipage}{0.48\textwidth}
 \centerline{\includegraphics[width=1.0\textwidth]{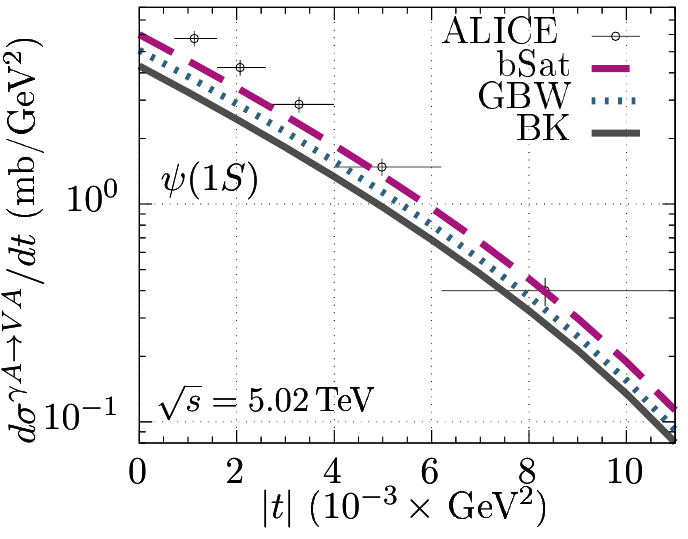}}
\end{minipage} \hfill
\begin{minipage}{0.48\textwidth}
 \centerline{\includegraphics[width=1.0\textwidth]{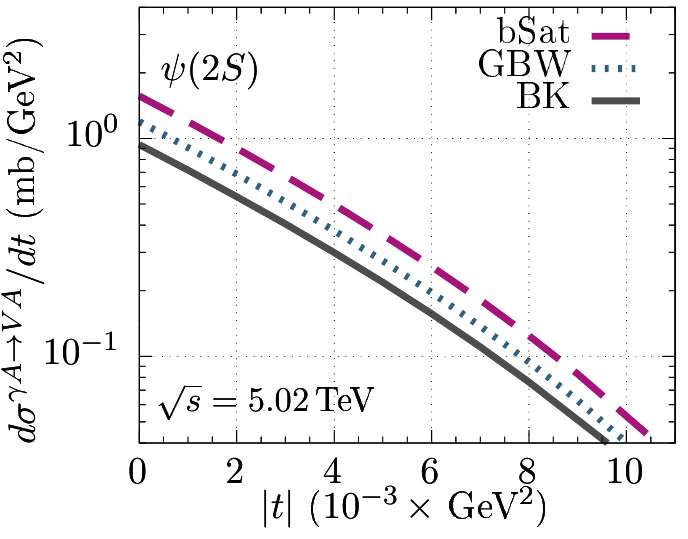}}
\end{minipage} \hfill
    \caption{Differential cross sections for the $\gamma Pb \rightarrow \psi(nS) Pb$
process as functions of $|t|$, with wavefunctions calculated using the Buchm\"{u}ller-Tye potential. The results using the BK and bSat dipole amplitudes and the purely phenomenological GBW dipole cross section \cite{GolecBiernat:1999qd} are compared with the recent ALICE data \cite{Acharya:2021bnz} for $\psi(1S)$.}
    \label{gammaA_cross_section_data}
\end{figure}

In Fig.~\ref{gammaA_cross_section_data} we compare the results for the differential cross section of $\psi(1S)$ state with the recent ALICE data \cite{Acharya:2021bnz} at $\sqrt{s} = 5.02 $ TeV. The models underestimate the $\psi(1S)$ data for photoproducion at small $|t|$, while getting closer to the available data at larger $|t|$. We also provide predictions for the differential cross section of $\psi(2S)$ at the same energy.

\begin{figure}[!h]
\begin{minipage}{0.48\textwidth}
 \centerline{\includegraphics[width=1.0\textwidth]{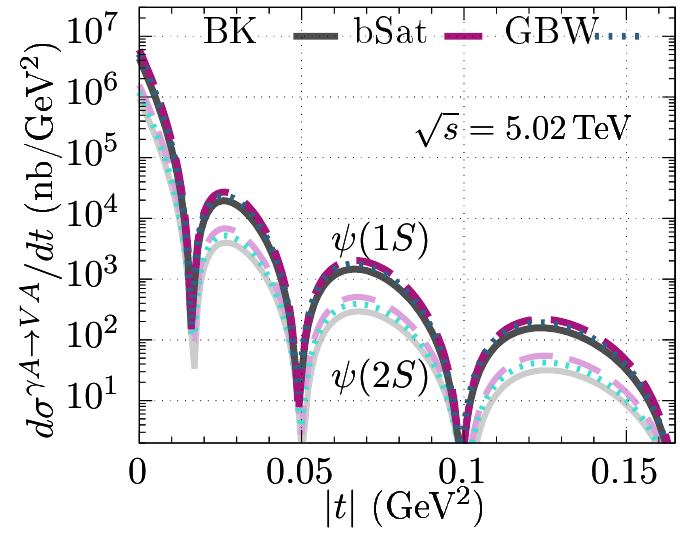}}
\end{minipage} \hfill
\begin{minipage}{0.48\textwidth}
 \centerline{\includegraphics[width=1.0\textwidth]{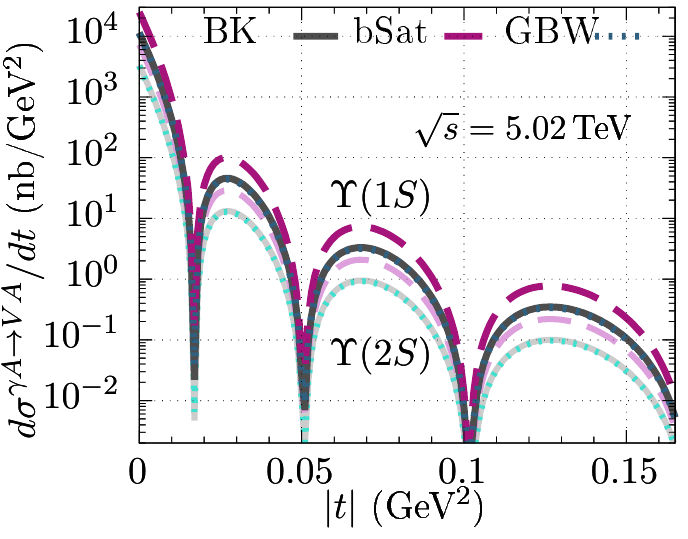}}
\end{minipage} \hfill
\caption{Predictions for the differential cross sections for the $\gamma Pb \rightarrow V Pb$
processes as functions of $|t|$, calculated with three dipole cross section models: the numerical solution of the BK equation for the dipole amplitude, the bSat model and the GBW parameterisation. The results for the production of $\psi$ states (left) 
 and $\Upsilon$ states (right) are shown. Both panels present the results at $y=0$ and with $\sqrt{s} = 5.02$TeV.
}
\label{gammaA_cross_section} 
\end{figure}

Predictions are presented in Fig.~\ref{gammaA_cross_section} for the differential cross section of the 
$\gamma Pb \rightarrow V Pb$ coherent photoproduction of $\psi$ states (left panel) 
and $\Upsilon$ states (right panel) at $\sqrt{s} = 5.02 $ TeV for a large range of $|t|$. We notice in this figure 
that the positions of the dips are almost the same for both $\psi(1S,2S)$ and $\Upsilon(1S,2S)$, 
which is caused by the destructive interference of individual scattering amplitudes of the nucleons 
of the target nucleus. 

\begin{figure}[!h]
\begin{minipage}{0.48\textwidth}
 \centerline{\includegraphics[width=1.0\textwidth]{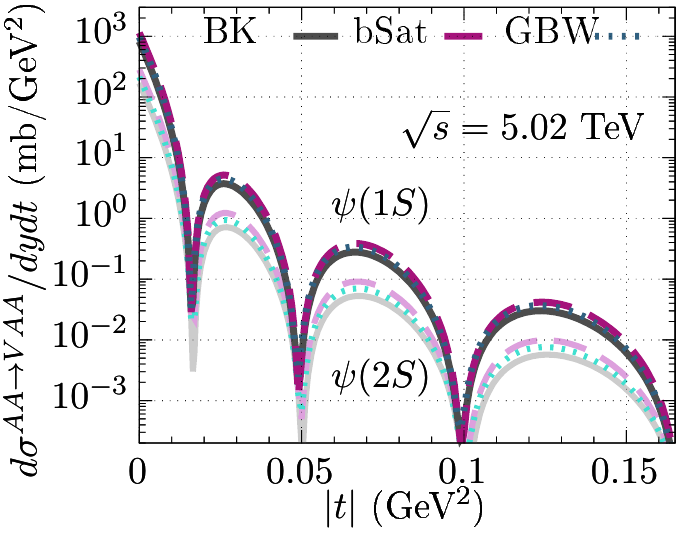}}
\end{minipage} \hfill
\begin{minipage}{0.48\textwidth}
 \centerline{\includegraphics[width=1.0\textwidth]{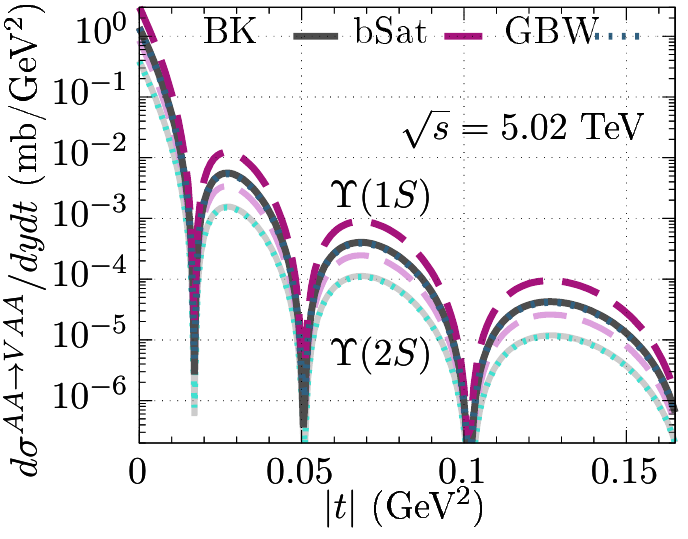}}
\end{minipage} \hfill
\caption{Predictions for the differential cross sections for $Pb Pb \rightarrow V Pb Pb$ processes as functions of
$|t|$. We show the production of $\psi$ states (left) and $\Upsilon$ states (right) at $\sqrt{s} = 5.02$ TeV and $y=0$.
}
\label{AA_cross_section} 
\end{figure}

In Fig.~\ref{AA_cross_section} we present a similar plot but for $AA \rightarrow V AA$ process in $AA$ UPCs for the LHC conditions (with lead nuclei), namely, at $\sqrt{s} = 5.02$ TeV. We also chose central ($y=0$) rapidity in order to maximise the corresponding differential cross sections and hence to increase the possibility of detection at the LHC. One obvious thing to mention is that these results have exactly the same shape as the ones in Fig.~\ref{gammaA_cross_section}, except that they are three orders of magnitude larger, which is caused by the photon flux. We have checked the validity of using the BK equation for a single nucleon plus the Glauber--Gribov approach together with some gluon shadowing by comparing our results to the ones given in Ref.~\cite{Bendova:2020hbb}, which  use the BK equation for the whole nucleus and found good agreement. 

\section{Conclusions}
\label{Sect:conclusions}

In this work, the impact-parameter $b$ dependent dipole model has been employed
for studies of differential (in momentum transfer squared $t$) observables of 
elastic (coherent) quarkonia photoproduction off proton and nuclear targets. 
In the treatment of quarkonia light-front wave functions, our work relies on 
the potential approach. Here, a radial-wave solution of the Schrödinger equation 
for a given interquark potential is first obtained in the $Q\bar Q$ rest frame and then 
boosted to the infinite momentum frame while the spin-dependent part of the wave function 
is computed by means of the Melosh transformation. We also incorporate the skewness effect 
in the partial dipole amplitude at the $\gamma p$ level, while in the nuclear case 
the dipole cross section for an elementary dipole scattering off a single nucleon 
has been multiplied by such a correction factor, and not the whole $\gamma A$ amplitude.
Besides, the gluon shadowing effect in photoproduction off a heavy nucleus target 
has been accounted for fully phenomenologically.

The use of the Buchm\"{u}ller-Tye potential together with the $b$-dependent solution of the Balitsky-Kovchegov (BK) equation for the dipole-target amplitude enables us to reproduce well the H1 data available from the HERA collider for differential $J/\psi$ photoproduction cross section with the proton target. The same setup has been used to make predictions for the $t$-dependent photoproduction $\gamma p \rightarrow Vp$ cross section of $\psi(2S)$, $\Upsilon(1S)$ and $\Upsilon(2S)$ vector mesons. A comparison with the corresponding results obtained by using the bsat model has revealed that the latter model gives a slightly larger cross section. This is the first calculation using a realistic potential model for the excited-state wave functions and based on the latest developments in the $b$-dependent BK equation as described above.

Furthermore, new predictions for the differential $\gamma Pb \rightarrow V Pb$ and 
$Pb Pb \rightarrow V Pb Pb$ cross sections at central rapidity have been 
reported for both ground and excited $\psi$ and $\Upsilon$ states. Our calculations are based upon the Glauber--Gribov picture of high energy scattering and included the gluon shadowing from a recent parametrization of nuclear PDFs. A single framework consistently combining these important elements has not yet been developed so far in the literature. The new data on the $J/\psi$ meson photoproduction recently published by the ALICE collaboration made it possible to test our approach. In this case, the bsat model provides a better description of the data than the other models, showing that there are still significant uncertainties in modelling of the $b$-dependent color dipole cross section.

Lastly, the results on the differential cross section for $\Upsilon(nS)$ photoproduction via $\gamma p$ interaction in $pPb$ collisions were compared with the existing CMS data, and again, the bsat model predictions appear to be closer to the data points than those of the BK model. Ultimately, these calculations are expected to be of large importance for further deeper investigations of the quarkonia coherent photoproduction mechanisms in ultraperipheral collisions in the future measurements at the LHC and at the electron--ion collider~\cite{Accardi:2012qut}.

\section*{Acknowledgments}

This work was supported by Fapesc, INCT-FNA (464898/2014-5), and CNPq (Brazil) for CH, EGdO, and HT. 
This study was financed in part by the Coordenação de Aperfeiçoamento de Pessoal de Nível 
Superior -- Brasil (CAPES) -- Finance Code 001. The work has been performed in 
the framework of COST Action CA15213 ``Theory of hot matter and relativistic heavy-ion collisions''
(THOR). R.P.~is supported in part by the Swedish Research Council grants, contract numbers
621-2013-4287 and 2016-05996, as well as by the European Research Council (ERC) under 
the European Union's Horizon 2020 research and innovation programme (grant agreement No 668679). 

\nocite{*}
\bibliography{bib}

\end{document}